\documentstyle[preprint,tighten,pra,aps]{revtex}
\begin{document}
\tightenlines
\draft

\title{One-loop Effective  Potential for a Fixed Charged Self-interacting 
Bosonic Model  at Finite Temperature with its
Related  Multiplicative Anomaly}

\author{Emilio Elizalde$^{1,}$\footnote{e-mail: eli@zeta.ecm.ub.es\rm},
Antonio Filippi$^{2,}$\footnote{e-mail: a.filippi@ic.ac.uk\rm},\\ 
Luciano Vanzo$^{3,}$\footnote{email: vanzo@science.unitn.it} and
Sergio Zerbini$^{3,}$\footnote{email: zerbini@science.unitn.it}
}
\address{ $^1$ Consejo Superior de Investigaciones
   Cient\'{\i}ficas, \\IEEC, Edifici Nexus 201,
   Gran Capit\`a 2-4, 08034 Barcelona, Spain\\ 
   and Departament ECM and IFAE, Facultat de F\'{\i}sica, \\ 
   Universitat de Barcelona, Diagonal 647, 08028 Barcelona, Spain}
\address{ $^2$ Theoretical Physics Group, Imperial
   College, \\Prince Consort Road, London SW7 2BZ, U.K.}
\address{ $^3$ Dipartimento di Fisica, Universit\`a di Trento 
   \\ and Istituto Nazionale di Fisica Nucleare 
   \\ Gruppo Collegato di Trento, Italia}

\maketitle\maketitle\begin{abstract}
The one-loop partition function for a charged self-interacting Bose gas at finite 
temperature in D-dimensional spacetime is evaluated within a path 
integral approach making use of zeta-function regularization. For D 
even, a new  additional vacuum term ---overlooked in all previous 
treatments and coming from the multiplicative anomaly related 
to functional determinants--- is found  and its dependence on 
the mass and chemical potential is obtained. The presence of the 
new term is   shown to be crucial for having the factorization invariance of 
the regularized partition function. In the non interacting 
case, the relativistic Bose-Einstein condensation is revisited. By 
means of a suitable charge renormalization, for $D=4$ the symmetry 
breaking phase is shown to be unaffected by the new term, 
which, however, gives actually rise to a non vanishing new contribution in the 
unbroken phase.  
\end{abstract}



\section{Introduction}
\label{Form}

The one-loop finite temperature effective potential with a non
vanishing chemical potential has been considered quite often during
the last years. The case of a free relativistic charged bosonic field
was investigated in Refs. \cite{kapu81-24-426,kapu89b,habe81-46-1497,habe82-23-1852}, while the self-interacting 
charged scalar field has been studied in
\cite{bens91-44-2480,kirs95}. Recently, the issue has been reconsidered
and critically analyzed in Ref. \cite{fili97u-90}, where use has been
made of zeta-function regularization in the calculations and a
different approach for including the chemical potential has been
proposed.

We would like to recall the importance of zeta-function
regularization, introduced in \cite{ray71-7-145,dowk76-13-3224,hawk77-55-133},
as a powerful tool to deal with the ambiguities (ultraviolet
divergences) present in  relativistic quantum field theories (see for
example \cite{eliz94b,byts96-266-1}). It permits to give a meaning
---in the sense of analytic continuation--- to the determinant of a
differential operator which, as the product of its eigenvalues, would
be formally divergent. For the sake of simplicity we shall here
restrict ourselves to scalar fields. In the case of a neutral scalar
field, we recall that the one-loop Euclidean partition function,
regularised by means of zeta-function techniques, reads
\cite{hawk77-55-133} \begin{eqnarray} \ln Z=-\frac{1}{2}\ln\det
\frac{L_D}{M^2} =\frac{1}{2}\zeta'(0|L_D)+\frac{1}{2}\zeta(0|L_D)\ln
M^2 \nonumber\:,\end{eqnarray} where $\zeta(s|L_D)=\mbox{Tr}\,   
L_D^{-s}$ is the zeta
function related to $L_D$, a second order elliptic differential
operator, $\zeta'(0|L_D)$ its derivative with respect to $s$, and
$M^2$ is a renormalization scale mass. The fact is used here that the
analytically continued zeta-function is generically regular at $s=0$,
and thus its derivative is well defined.

In this paper, we shall take full advantage of a rigorous use of this 
regularization  in investigating the finite temperature effects in
presence of a non vanishing chemical potential. In order to illustrate 
the new issue we shall be dealing with, we start with the
 partition function for a free charged field in
${\hbox{{\rm I}\kern-.2em\hbox{\rm R}}}^D$, described by two real components $\phi_i$. The Euclidean action is \begin{eqnarray}
S=\int  dx^D  \left[ \phi_i\left(-\Delta_D+m^2 \right)
\phi_i \right]
\:,\label{ea}\end{eqnarray} where $\phi^2=\phi_k \phi_k$ is $O(2)$
invariant. 

As is well known, in order to study finite temperature effects, 
one observes that the
grand canonical partition function may be written under the form (we shall call
this method of including the chemical potential, as discussed in Ref.
\cite{fili97u-90},  method I) \begin{eqnarray}
Z_{\beta}(\mu)=\mbox{Tr}\,   e^{-\beta (H-\mu Q)}
\:,\label{gp}\end{eqnarray} in which $H$ is the Hamiltonian, $Q$ the
conserved charge operator and $\beta$  the inverse of the equilibrium
temperature. Making use of a path integral representation  and
integrating over the momenta, one arrives at the following recipe
\cite{kapu81-24-426,kapu89b,bens91-44-2480,fili97u-90,habe82-25-502}:
(i) compactify the imaginary time $\tau$ in the interval $[0, \beta]$,
(ii) assume a periodic boundary condition in $\beta$, and (iii)
include the chemical potential $\mu$ by adding to the action the term
\begin{eqnarray} i\mu e \epsilon_{ik}\phi_i \partial_\tau
\phi_k-\frac{1}{2}e^2 \mu^2 \phi^2 \:,\label{cp}\end{eqnarray} where
$e$ is the elementary charge. As a result, the  partition
function reads \begin{eqnarray} Z_{\beta}(\mu)=
\int_{\phi(\tau)=\phi(\tau+\beta)}[d \phi_i]
e^{-\frac{1}{2}\int_0^\beta d\tau \int d^Nx \phi_iA_{ij}\phi_j}
\:,\label{}\end{eqnarray} where $D=N+1$  and $A$ is given by \begin{eqnarray}
A_{ij}=L_{ij}+ 2e \mu \epsilon_{ij} \sqrt{L_\tau}
\:,\label{sd}\end{eqnarray} with \begin{eqnarray} L_{ij}=\left(
L_\tau+ L_N-e^2 \mu^2
\right)\delta_{ij},\,\,\, L_N=-\Delta_N+m^2 \:,\label{nm}\end{eqnarray}
in which $\Delta_N$ is the Laplace operator on $ {\hbox{{\rm
I}\kern-.2em\hbox{\rm R}}}^N$ (continuous spectrum $\vec k^2$) and
$L_\tau=-\partial^2_\tau$ (discrete spectrum $\omega_n^2=\frac{4\pi^2
n^2}{\beta^2}$), the Laplace operator on $S^1$. Thus, one is actually
dealing with a  matrix-valued elliptic differential non self-adjoint
operator acting on scalar fields in $S^1 \times  
{\hbox{{\rm I}\kern-.2em\hbox{\rm R}}}^N$. 
In this case,  the partition function
may be written as \cite{bens91-44-2480} \begin{eqnarray} \ln
Z_{\beta}(\mu)=-\frac{1}{2}\ln\det \left\| \frac{A_{ik}}{M^2}\right\|
=-\frac{1}{2}\ln\det \left[ \frac{L_+}{M^2}\frac{L_-}{M^2}\right],
\label{o2} \end{eqnarray} where \begin{eqnarray} L_\pm=L_\tau +
L_N+e^2 \mu^2 \pm 2 e \mu  (L_N)^{\frac{1}{2}} \:.\label{bbd}\end{eqnarray}
Another possible factorization (see, for example, \cite{byts96-266-1}) is 
\begin{eqnarray}
K_\pm=L_N+L_\tau-e^2 \mu^2 \pm 2 i e \mu  (L_\tau)^{\frac{1}{2}} 
\:.\label{z}\end{eqnarray} Of
course we have $L_+L_-=K_+K_-$ and, in both cases,  one is dealing
with a couple of pseudo-differential operators ($\Psi$DOs), $L_+$ and
$L_-$ being also formally self-adjoint.

As is clear, in any case  the product of two elliptic $\Psi$DOs, say $A_+$
and $A_-$, appears. It is known that, in general, the zeta-function
regularized determinants do {\it not} satisfy the relation $\det
(AB)=\det A \det B$. In fact, in general,  there appears the so-called
multiplicative anomaly \cite{kass89-177-199,kont95b}. In terms of
$F(A,B)\equiv \det (AB)/(\det A \det B)$ \cite{kont95b}, it is defined
as: \begin{eqnarray} a_D(A,B)=\ln F(A,B)=\ln \det (AB)-\ln \det
(A)-\ln \det (B) \:,\label{ma}\end{eqnarray} in which the determinants
of the two elliptic operators, $A$ and $B$, are assumed to be defined
(i.e. regularized) by means of the zeta function \cite{ray71-7-145}.
Thus, the partition function, chosen a factorization $ A_+A_-$, is
given by \begin{eqnarray} \ln Z_{\beta}(A_+,A_-)&=&-\frac{1}{2}\ln\det
\left\| \frac{A_{ik}}{M^2}\right\| =-\frac{1}{2} \ln\det \left[
\frac{A_+}{M^2}\frac{A_-}{M^2}\right]\label{pfc} \\ 
&=&\frac{1}{2}\zeta'(0|A_+)+\frac{1}{2} \zeta'(0|A_-)+\frac{\ln
M^2}{2} \left[ \zeta(0|A_+)+\zeta(0|A_-) \right]
-\frac{1}{2}a_D(A_+,A_-) \:.\nonumber\end{eqnarray} Here one can see
the crucial  role of the multiplicative anomaly: as, in the
factorization, different operators may enter, the multiplicative
anomaly is necessary in order to have the same regularized partition
function in both cases, namely $\ln Z_{\beta}(A_+,A_-)=\ln
Z_{\beta}(B_+,B_-)$, which is an obvious physical requirement. (What
it is very easy to see in general is that det $A_+$ det $A_- \neq$ det
$B_+$ det $B_-$.)

With regards to the multiplicative anomaly, an important result is
that it  can be expressed by means of the  non-commu\-ta\-ti\-ve
residue associated with a classical $\Psi$DO, known as the Wodzicki
residue \cite{wodz87b}.

In the limit of zero temperature and vanishing
chemical potential, the two elliptic $\Psi$DOs $L_+$ and $L_-$ reduce
to Laplace-type second order differential operators with constant
potential  terms defined in $ {\hbox{{\rm I}\kern-.2em\hbox{\rm
R}}}^4$. In this case, the multiplicative anomaly has been computed in
Ref. \cite{eliz97u-394}.

One could argue that the presence of a multiplicative anomaly is
strictly linked with the zeta-function regularization employed and,
thus, that it might be an artifact of it. In the Appendix we will
prove that the multiplicative anomaly is present indeed in a large
class of regularizations of functional determinants appearing in the
one-loop effective action.

The contents of the paper are the following. In Sec. 2 we reconsider
the free case making use of zeta-function regularization and we carry
out a comparison of the two possible factorizations. In Sec. 3, the 
self-interacting case is presented in the one-loop approximation. In Sec. 4 we
briefly introduce the Wodzicki residue and present a proof of the
Wodzicki formula expressing the multiplicative anomaly in terms of the
corresponding non-commutative residue of a suitable classical $\Psi$DO.  In Sec.
5, the results of Wodzicki are used in the computation of the
multiplicative anomaly for the interacting $O(2)$ model. In Sec. 6, the
Bose-Einstein condensation phenomenon is discussed in the
non-interacting case. Some final remarks are presented in the
concluding section. In the Appendix, the presence of the
multiplicative anomaly in a very large class of functional determinant
regularizations is proven.

\section{Chemical potential in the non-interacting case revisited}

We shall treat here the non-interacting
bosonic model at finite temperature with a non vanishing chemical
potential. This situation has been extensively investigated in the
past. We will reconsider it here making rigorous use of the
zeta-function regularization procedure, what will allow us to clarify 
some subtle points overlooked in previous studies.

Again, it is convenient to work in $S^1 \times  {\hbox{{\rm
I}\kern-.2em\hbox{\rm R}}}^N$, with $D=N+1$ in order to exploit the
role of the dimension of the space-time. As discussed in the 
Introduction, here  one has to deal with the functional determinant of 
the product of the two operators \begin{eqnarray} L_\pm=L_\tau+\left( \sqrt L_N \pm e \mu \right)^2
\:,\label{bbd1}\end{eqnarray} while a second factorization is obtained
in terms of the operators \begin{eqnarray} K_\pm=L_N+\left( \sqrt L_\tau \pm i e \mu
\right)^2 \:,\label{z1}\end{eqnarray} with $L_N=-\Delta_N+m^2$.

Let us start with the latter one, since it renders calculations easier.
The zeta function can be defined, for a sufficiently large real part 
of $s$, by means of the Mellin transform of
the related heat operator trace, namely 
\begin{eqnarray}
\zeta(s|K_\pm)=\frac{1}{\Gamma(s)}\int_0^\infty dt\
 t^{s-1}\mbox{Tr}\,   \exp
\left( -tK_\pm\right)
\:,\label{zetam}\end{eqnarray}
where
$\mbox{Tr}\,   \exp
-tK_\pm$ is given by \begin{eqnarray} \mbox{Tr}\,   e^{
-tK_\pm}= \mbox{Tr}\,   e^{ -t L_N}\sum_n  e^{ -t(\omega_n
\pm i e \mu)^2} \:.\label{hk}\end{eqnarray} For notational simplicity,
we  write \begin{eqnarray} \mbox{Tr}\,   e^{ -t L_N}=\sum_i
e^{ -t \lambda_i}=\frac{V_N}{(4\pi t)^{\frac{N}{2}}}e^{-tm^2}
\:,\label{90}\end{eqnarray} with $\lambda_i$ the eigenvalues of the
operator $L_N$, namely $\vec k^2+m^2$. Thus, via Mellin transform, one
has the standard result \begin{eqnarray} \zeta(z|L_N)=\frac{ V_N}{(4
\pi)^{\frac{N}{2}}}\frac{\Gamma(z-\frac{N}{2})}{\Gamma (z)} m^{N-2z}
\:.\label{ze}\end{eqnarray} The Poisson-Jacobi resummation formula
gives \begin{eqnarray} \sum_n  e^{ -t(\omega_n \pm
i\mu)^2}=\frac{\beta}{(4 \pi t)^{\frac{1}{2}}} \left(
1+2\sum_{n=1}^\infty \cosh (n e \mu \beta) e^{-\frac{n^2 \beta^2}{4t}}
\right) \:.\label{kt2}\end{eqnarray} As a result, even though $K_+
\neq K_-$, one gets \begin{eqnarray} \mbox{Tr}\,   e^{
-tK_+}&=&\mbox{Tr}\,   e^{ -tK_-} \nonumber \\ 
&=&\mbox{Tr}\,   e^{ -t L_N}\frac{\beta}{(4 \pi t)^{1/2}}
\left( 1+2\sum_{n=1}^\infty \cosh (n e \mu \beta) e^{-\frac{n^2
\beta^2}{4t}} \right) \:,\label{kt3}\end{eqnarray} and
$\zeta(s|K_+)=\zeta(s|K_-)$. Therefore, making use of 
Eq.~(\ref{zetam}), Eq.~(\ref{hk}), Eq.~(\ref{kt2}) and of the well 
known integral 
representation of the MacDonald function $K_s(x)$,
\begin{eqnarray}
K_s(\sqrt b x)=\frac{1}{2} \left(\frac{x}{2 \sqrt b} \right)^{s}\int_0^\infty  
e^{-b t-\frac{x^2}{4t}} t^{-s-1} dt
\:,\label{mdf}\end{eqnarray}
one has
\begin{eqnarray}
\frac{1}{2}\left[ \zeta(s|K_+)+\zeta(s|K_-) \right] &=&\frac{\beta
V_N}{(4 \pi)^{\frac{D}{2}}}\frac{\Gamma(s-\frac{D}{2})}{\Gamma (s)}
m^{D-2s}\nonumber \\ &+&\frac{2^{3/2-s}\beta}{\sqrt \pi \Gamma(s)}
\sum_{n=1}^\infty \cosh (n e\mu \beta) (n\beta)^{s-\frac{1}{2}} \sum_i
\lambda_i^{\frac{1-2s}{4}} K_{s-\frac{1}{2}}(n\beta \sqrt \lambda_i)
\:.\label{kt4}\end{eqnarray} The second term in this equation represents
the statistical sum contribution, which depends non-trivially on the
temperature and chemical potential and can  be calculated in arbitrary
dimension. Near $s=0$ it can be written as $s S(\beta,\mu)+O(s^2)$,
with \begin{eqnarray} S(\beta,\mu)=-\sum_i \left[ \ln \left(
1-e^{-\beta(\sqrt \lambda_i+ e \mu)} \right)+\ln \left(
1-e^{-\beta(\sqrt \lambda_i-e \mu)} \right) \right]
\:.\label{ss}\end{eqnarray} Taking Eq.~(\ref{90}) into account, this
contribution can also be rewritten as follows \begin{eqnarray}
S(\beta,\mu)=\frac{4V_N\beta}{(4 \pi)^{\frac{D}{2}} }
\sum_{n=1}^\infty \cosh (n e \mu \beta) (n\beta)^{-\frac{D}{2}}
(2m)^{\frac{D}{2}} K_{\frac{D}{2}}(n\beta m)
\:.\label{ss1}\end{eqnarray} It should be noted that this statistical
sum contribution is a series involving MacDonald functions. This
series is always convergent  in any $D >1$ space-time dimension and
for any $\beta$ and $e |\mu| \leq m$, even  in the critical limit $ e
|\mu_c|=m$, $m$ being the lowest eigenvalue in the spectrum of $L_N$ .
This follows from the asymptotics of $K_\nu (z)$ for large $z$:
\begin{eqnarray} K_\nu(z) \simeq \left( \frac{\pi}{2z}
\right)^{\frac{1}{2}} e^{-z} \left[ 1+O(\frac{1}{z})\right] \:.
\label{md} \end{eqnarray} 

Coming back to Eq.~(\ref{kt4}),  the first term (the vacuum
contribution)  has to be considered, as usual, for $D$ odd and $D$
even separately. For $D$  odd, we have \begin{eqnarray}
\zeta(0|K_+)+\zeta(0|K_-)=0 \:,\label{odd1}\end{eqnarray} and
\begin{eqnarray} \frac{1}{2}\left(
\zeta'(0|K_+)+\zeta'(0|K_-)\right)=\frac{\beta V_N}{(4
\pi)^{\frac{D}{2}}}\Gamma(-\frac{D}{2}) m^{D}+S(\beta,\mu)
\:.\label{odd2}\end{eqnarray} For $D$ even, we shall restrict
ourselves to the two cases $D=2$ and $D=4$. We obtain \begin{eqnarray}
\zeta(0|K_+)\ln M^2+\zeta'(0|K_+)&=&-\frac{\beta V_1
m^2}{4\pi}\left(\ln \frac{m^2}{M^2}-1 \right)+S(\beta,\mu)\,,\,\, D=2
\:,\label{d2}\\ \zeta(0|K_+)\ln M^2+\zeta'(0|K_+)&=&-\frac{\beta V_3
m^4}{ 32\pi^2}\left(\ln \frac{m^2}{M^2}-\frac{3}{2}
\right)+S(\beta,\mu)\,,\,\, D=4 \:.\label{d4}\end{eqnarray}

Let us now consider the first factorization. \begin{eqnarray}
\mbox{Tr}\,   e^{ -tL_\pm}=\mbox{Tr}\,   e^{ -t
(\sqrt L_N \pm e\mu)^2}\sum_n  e^{ -t\omega_n^2}
\:,\label{19}\end{eqnarray} with \begin{eqnarray} \mbox{
Tr}\, e^{ -t (\sqrt L_N \pm e\mu)^2}=\sum_i e^{ -t (\sqrt
\lambda_i \pm e\mu)^2} \:.\label{901}\end{eqnarray} Again, Poisson
resummation gives \begin{eqnarray} \mbox{Tr}\,   e^{
-tL_\pm}=\mbox{Tr}\,   e^{ -t (\sqrt L_N \pm e
\mu)^2}\frac{\beta}{(4 \pi t)^{1/2}} \left( 1+2\sum_{n=1}^\infty 
e^{-\frac{n^2 \beta^2}{4t}} \right) \:,\label{10}\end{eqnarray} and,
as a consequence, \begin{eqnarray} \zeta(s|L_\pm)&=&\frac{\beta}{2
\sqrt
\pi}\frac{\Gamma(s-\frac{1}{2})}{\Gamma(s)}\zeta(s-\frac{1}{2}|(\sqrt
L_N \pm e\mu)^2) \nonumber \\ &-&2 s  \sum_i \ln \left(
1-e^{-\beta(\sqrt \lambda_i \pm e\mu)} \right)+O(s^2)
\:.\label{98}\end{eqnarray} Now, for $e|\mu| < m$, the binomial
theorem yields \begin{eqnarray} \zeta(z|(\sqrt L_N +
e\mu)^2)+\zeta(z|(\sqrt L_N - e\mu)^2)&=& 2\zeta(z|L_N)\nonumber \\ 
&+&2\sum_{r=1}^\infty
\frac{\Gamma(2z+2r)}{(2r)!\Gamma(2z)}(e\mu)^{2r}\zeta(z+r|L_N)
\:,\label{ioo}\end{eqnarray} and this leads to \begin{eqnarray}
\frac{1}{2} \left( \zeta(s|L_+)+\zeta(s|L_-) \right)&=&\frac{\beta V_N
m^{D-2s}} {( 4\pi)^{\frac{D}{2}}} \left\{
\frac{\Gamma(s-\frac{D}{2})}{\Gamma(s)} \right. \label{mn} \\ &+&
\left.  \sum_{r=1}^\infty
\frac{\Gamma(2s+2r-1)\Gamma(s-\frac{1}{2})}{(2r)!\Gamma(s)\Gamma(s-\frac{1}{2}+r)\Gamma(2s-1)}\left(
\frac{e \mu}{m}\right)^{2r} \Gamma(s+r-\frac{D}{2})\right\} \nonumber
\\ &-&sS(\beta,\mu)+O(s^2) \:.\nonumber \end{eqnarray} As a result,
for $D$ odd, taking the derivative with respect to $s$ and the limit $
s \to 0$, the two factorizations give \begin{eqnarray} \ln
Z_{\beta,\mu}(L_+,L_-)=\frac{\beta V_N m^D} {(4
\pi)^{\frac{D}{2}}}\Gamma(-\frac{D}{2})+S(\beta,\mu)-\frac{1}{2}a_D(L_+,L_-)
\:,\label{oddd}\end{eqnarray} and \begin{eqnarray} \ln
Z_{\beta,\mu}(K_+,K_-)=\frac{\beta V_N m^D} {(4
\pi)^{\frac{D}{2}}}\Gamma(-\frac{D}{2})+S(\beta,\mu)-\frac{1}{2}a_D(K_+,K_-)
\:,\label{oddd1}\end{eqnarray} respectively. It should be noted that
one gets $\ln Z_{\beta,\mu}(L_+,L_-)=\ln Z_{\beta,\mu}(K_+,K_-)$ and
thus the standard textbook result  as soon as one is able to prove
that the two multiplicative anomalies are vanishing for $D$ odd. We
can anticipate that this indeed happens.

For $D$ even, the situation is different because, within the first
factorization, the vacuum sector depends explicitly on the chemical
potential $\mu$. In fact,  for example, one has for $ D=2$
\begin{eqnarray} \frac{1}{2}\left\{ \zeta'(0|L_+)+\zeta'(0|L_-)+\left[
\zeta(0|L_+)+\zeta(0|L_+)\right] \ln M^2 \right\} &=&\frac{\beta V_1
}{\pi}\left[ m^2 (\ln \frac{m^2}{M^2}-1) \right] \nonumber \\ 
&+&\frac{\beta V_1}{2\pi}e^2 \mu^2 +S(\beta,\mu) \:,
\label{e2}\end{eqnarray} and for $ D=4$ \begin{eqnarray} &&
\hspace{-1cm} \frac{1}{2}\left\{ \zeta'(0|L_+)+\zeta'(0|L_-)+ \left[
\zeta(0|L_+)+\zeta(0|L_+)\right] \ln M^2 \right\} \nonumber \\ 
&=&\frac{\beta V_3 }{32\pi^2}\left[ m^4 (\ln
\frac{m^2}{M^2}-3/2)\right] + \frac{\beta V_3 }{8\pi^2} 2\left[
\frac{e^4\mu^4}{3}-e^2\mu^2 m^2  \right]+S(\beta,\mu)
\:,\label{e4}\end{eqnarray} As a result, for $D=2$ the two
factorizations give \begin{eqnarray} \ln
Z_{\beta,\mu}(L_+,L_-)=\frac{\beta V_1 m^2} { 4\pi}(\ln
\frac{m^2}{M^2}-1) +S(\beta,\mu)+\frac{V_1 \beta}{2\pi}
e^2\mu^2-\frac{1}{2}a_2(L_+,L_-) \:,\label{22}\end{eqnarray} and
\begin{eqnarray} \ln Z_{\beta,\mu}(K_+,K_-)=\frac{\beta V_1 m^2} {
4\pi}(\ln \frac{m^2}{M^2}-1)+S(\beta,\mu)-\frac{1}{2}a_2(K_+,K_-)
\:,\label{222}\end{eqnarray} respectively, while for $D=4$ one has
\begin{eqnarray} \ln Z_{\beta,\mu}(L_+,L_-)&=&\frac{\beta V_3
}{32\pi^2}\left[ m^4 (\ln \frac{m^2}{M^2}-3/2) \right]+ \frac{\beta
V_3 }{8\pi^2} \left( \frac{e^4\mu^4}{3}-e^2\mu^2 m^2 \right)\nonumber
\\ &+&S(\beta,\mu)-\frac{1}{2}a_4(L_+,L_-) \:,\label{44}\end{eqnarray}
and \begin{eqnarray} \ln Z_{\beta,\mu}(K_+,K_-)=\frac{\beta V_3
}{32\pi^2}\left[ m^4 (\ln \frac{m^2}{M^2}-3/2)
\right]+S(\beta,\mu)-\frac{1}{2}a_4(K_+,K_-)
\:.\label{441}\end{eqnarray} In general for $D$ even, one gets
\begin{eqnarray} \ln Z_{\beta,\mu}(L_+,L_-)&=&-\beta V_N {\cal
E}_V(m,M)+\beta V_N {\cal E}_D(m,\mu)
+S(\beta,\mu)-\frac{1}{2}a_D(L_+,L_-) \:,\label{44b}\end{eqnarray} and
\begin{eqnarray} \ln Z_{\beta,\mu}(K_+,K_-)=-\beta V_N {\cal
E}_V(m,M)+S(\beta,\mu)-\frac{1}{2}a_D(K_+,K_-)
\:,\label{441b}\end{eqnarray} where ${\cal E}_V(m,M)$ is the naive
vacuum energy density, $D=2Q$  and \begin{eqnarray} {\cal E}_D(m,\mu)=
\sum_{r=1}^{Q}c_{Q,r} (e \mu)^{2r}m^{D-2r}
\:.\label{calli}\end{eqnarray} Here the $c_{Q,r}$ are computable
coefficients, given by \begin{eqnarray}
c_{Q,r}=\frac{\Gamma(2r-1)}{\Gamma(2r+1)\, \pi^{\frac{N}{2}}}
\frac{2^{1-N}}{\Gamma(r-\frac{1}{2})}\frac{(-1)^{Q-r}}{(Q-r)!}
\:.\label{vc}\end{eqnarray} In the cases $D=2$, $D=4$ and $D=6$ one
obtains \begin{eqnarray} {\cal E}_2(m,\mu)&=&\frac{1}{2
\pi}e^2\mu^2\,, \qquad {\cal E}_4(m,\mu)=-\frac{1}{8\pi^2} \left[
e^2\mu^2\left( m^2-\frac{e^2\mu^2}{3}\right) \right] \nonumber \\ 
{\cal E}_6(m,\mu)&=&-\frac{1}{16\pi^3} \left[ e^2\mu^2\left(
-\frac{1}{4}m^4+\frac{e^2\mu^2 m^2}{6}-\frac{2 e^4 \mu^4 }{45}\right)
\right] \:.\label{ex}\end{eqnarray}

\section{The interacting case in the one-loop approximation}

In this section, we will review the interacting charged boson model 
and we will compute, within the one-loop approximation, the corresponding 
operators appearing in the two possible factorizations of the partition 
function.
 
The Euclidean action for a self-interacting charged 
field, again  described by two real components $\phi_i$, is 
\begin{eqnarray}
S=\int  dx^D  \left[ \phi_i\left(-\Delta_D+m^2 \right)
\phi_i+\lambda_D(\phi^2)^{\frac{D}{D-2}} \right]
\:,\label{ea1}\end{eqnarray} where $\phi^2=\phi_k \phi_k$ is $O(2)$
invariant and $\lambda_D=\frac{\lambda}{D!}$ is a dimensionless
coupling constant. In the important case $D=4$ one has \begin{eqnarray}
S=\int  dx^4  \left[ \phi_i\left(-\Delta+m^2 \right)
\phi_i+\frac{\lambda}{4!}(\phi^2)^2 \right]
\:.\label{ea11}\end{eqnarray}

Finite temperature effects with non vanishing chemical potential are 
accounted for in the following one-loop partition function 
\cite{kapu81-24-426,kapu89b,habe82-25-502,bens91-44-2480}
 \begin{eqnarray} Z_{\beta}(\mu)=
e^{-S_0}\int_{\phi(\tau)=\phi(\tau+\beta)}[d \phi_i]
e^{-\frac{1}{2}\int_0^\beta d\tau \int d^Nx \phi_iA_{ij}\phi_j}
\:,\label{bbbbo}\end{eqnarray} where \begin{eqnarray} S_0=\beta
V_N \left[ \frac{1}{2}(m^2-e^2 \mu^2)\Phi^2+\lambda_D
(\Phi^2)^{\frac{D}{D-2}}\right] \:,\label{s00}\end{eqnarray} $\Phi$
being the background field, assumed to be constant, and $A$ the
Euclidean small disturbances operator, given by \begin{eqnarray}
A_{ij}=L_{ij}+ 2e \mu \epsilon_{ij} \sqrt{L_\tau}+\frac{8
D\lambda_D}{(D-2)^2}(\Phi^2)^{\frac{4-D}{D-2}} \Phi_i
\Phi_j\:,\label{sd1}\end{eqnarray} with \begin{eqnarray} L_{ij}=\left(
L_\tau+ L_N-e^2 \mu^2+\frac{2D\lambda_D}{D-2}(\Phi^2)^{\frac{2}{D-2}}
\right)\delta_{ij} \:.\label{nm1}\end{eqnarray}
Again, one is 
dealing with a  matrix-valued, elliptic, differential, non self-adjoint
operator acting on scalar fields in $S^1 \times  {\hbox{{\rm
I}\kern-.2em\hbox{\rm R}}}^N$. As a consequence,  the partition function
reads \cite{bens91-44-2480} \begin{eqnarray} \ln
Z_{\beta}(\mu)=-\frac{1}{2}\ln\det \left\| \frac{A_{ik}}{M^2}\right\|
=-\frac{1}{2}\ln\det \left[ \frac{L_+}{M^2}\frac{L_-}{M^2}\right]
\label{o21} \end{eqnarray} where \begin{eqnarray} L_\pm=L_\tau +
L_N+e^2 \mu^2+ h_D \pm \left[ \frac{4}{D^2} h_D^2+4e^2 \mu^2 (L_N+h_D)
\right]^{\frac{1}{2}} \:,\label{bbd11}\end{eqnarray} with
\begin{eqnarray} h_D=\frac{2D^2\lambda_D}{(D-2)^2}
(\Phi^2)^{\frac{2}{D-2}}\,,\qquad h_4=\frac{\lambda}{3} \Phi^2\,.
\end{eqnarray} 
The other factorization is 
\begin{eqnarray}
K_\pm=L_N+L_\tau-e^2\mu^2+h_D \pm \left( \frac{4}{D^2} h_D^2-4e^2
\mu^2 L_\tau \right)^{\frac{1}{2}} \:,\label{z12}\end{eqnarray} and,
again, we have $L_+L_-=K_+K_-$ and we have found the couples of $\Psi$DOs
which enter in the multiplicative anomaly. The necessity of a general formula
for computing the multiplicative
anomalies is clear. This issue will be discussed in the next section.

\section{The Wodzicki residue and the multiplicative anomaly formula}

For the reader's convenience, we will review in this section the
necessary information concerning the Wodzicki residue  \cite{wodz87b}
(see \cite{kass89-177-199} and the references to Wodzicki 
quoted therein) that will be used in the rest of the paper. Let us
consider a $D$-dimensional smooth compact manifold without boundary
$M_D$ and a (classical) $\Psi$DO, $A$,  acting on sections of vector
bundles on $M_D$. To any classical $\Psi$DO, $A$, it corresponds a complete
symbol $A(x,k)=e^{-ikx}Ae^{ikx}$, such that, modulo infinitely smoothing operators, one
has \begin{eqnarray} (A f)(x)\sim
\int_{{\hbox{{\rm I}\kern-.2em\hbox{\rm R}}}^D}\frac{dk}{(2\pi)^{D}}
\int_{{\hbox{{\rm I}\kern-.2em\hbox{\rm R}}}^D}dy \ e^{i(x-y)k}A(x,k)f(y)
\:.\label{sy}\end{eqnarray} The complete symbol admits an asymptotic
expansion for $|k| \to \infty$, given by the series \begin{eqnarray} A(x,k)\sim
\sum_j A_{a-j}(x,k) \:,\label{sy1}\end{eqnarray} and the coefficients 
(their number is infinite) fulfill the
homogeneity property $ A_{a-j}(x,tk)= t^{a-j}A_{a-j}(x,k)$, for $t>0$.
The number $a$ is called the order of $A$. For example, in the case of 
differential operators, the complete symbol can be obtained by the 
substitution $ \partial_\mu \rightarrow i k_\mu$ and it has a finite number of 
coefficients and the series stops at $A_0(x,k)$, being a polynomial in 
$k$ of order $a$.

Now let us introduce the notion of non-commutative residue of a 
classical $\Psi$DO $A$ of order $a$. 
If $P$ is an elliptic operator of order $p>a$, according to
Wodzicki one can construct the  $\Psi$DO $P_A(u)$, $\mbox{ord} P_A(u)=p$
\begin{eqnarray}
P_A(u)=P+uA
\:,\label{res1}\end{eqnarray}  
and its related zeta-function
\begin{eqnarray}
\zeta(s|P_A(u))=\mbox{Tr}\,   \left( P+uA \right)^{-s} 
\:.\label{r2}\end{eqnarray}
This zeta-function has a meromorphic analytical continuation, which 
can be determined by the standard method starting from the short $t$ 
asymptotics of $\mbox{Tr}\,   \exp{-t P_A(u)}$. Since  $P_A(u)$ 
is a  $\Psi$DO of order $p$, the heat-kernel asymptotics  reads \cite{kass89-177-199} 
\begin{eqnarray}
\mbox{Tr}\,    e^{-t P_A(u)}\simeq \sum_{j=0}^\infty \alpha_j(u) t^{\frac{j-D}{p}}
+\sum_{k=1}^\infty\beta_k(u)t^k \ln t
\:.\label{r3}\end{eqnarray}
It should be noted the presence  of logarithmic terms in this 
asymptotic expansion, absent if one is dealing with a differential 
elliptic operator. Taking the derivative with respect to the parameter 
$u$, one also gets the meromorphic structure of $\lim_{u \to 
0}\frac{d}{du} \zeta(s|P_A(u))$. By definition, the non-commutative 
residue of $A$ is 
\begin{eqnarray}
\mbox{res}(A)= p \mbox{Res}\,   \left[ \lim_{u \to 0}\frac{d}{du} \zeta(s|P_A(u))\right]_{s=-1}
\:,\label{r4}\end{eqnarray} 
where $\mbox{Res}\,   $ is the usual Cauchy residue.
It is possible to show that this definition is independent of the
elliptic operator $P$ and that the trace of the operator  $AP^{-s}$ exists
and admits a meromorphic continuation to the whole complex plane, with a
simple pole at $s=0$. Its Cauchy residue  at $s=0$ is proportional
to non-commutative  residue of $A$:
\begin{eqnarray}
\mbox{res} (A)=p \mbox{Res}\,  _{s=0} \mbox{Tr}\,   (AP^{-s})
\:.\label{wod1}\end{eqnarray}
Strictly related to the latter
result is the following one, involving the short-$t$ asymptotic expansion
\begin{eqnarray}
\mbox{Tr}\,   (A e^{-tP})\simeq \sum_j c_j t^{\frac{j-D}{p}-1}-\frac{\mbox{res}
(A)}{p} \ln t+O(t \ln t)
\:.\label{wod11}\end{eqnarray}
Thus, the Wodzicki residue of the  $\Psi$DO $A$ can be read off
from the above asymptotic expansion, selecting the coefficient
proportional to $\ln t$.
 Furthermore, it is possible to show that  $\mbox{res} (A)$ is
linear with respect to $A$ and that it possesses the  important
property of being the unique trace on the algebra of
the classical $\Psi$DOs, namely, one has $\mbox{res}(AB)=\mbox{res}(BA)$. 

 Wodzicki has also obtained a local form of the
non-commutative residue, which has the fundamental consequence of
characterizing it through a scalar density. This density  can be
integrated to yield the Wodzicki residue, namely
\begin{eqnarray}
\mbox{res}(A)=\int_{M_D}\frac{dx}{(2\pi)^{D}}\int_{|k|=1}A_{-D}(x,k)dk
\:.\label{wod2}\end{eqnarray}
Here the homogeneous component $A_{-D}(x,k)$ of order $-D$ of the complete symbol appears.
The above result  leads to  $\mbox{res} (A)=0$
when $A$ is an elliptic  differential operator.

Now let us discuss  the multiplicative anomaly formula, due to
Wodzicki.  A more general expression has been derived in
\cite{kont95b}. Consider two invertible, self-adjoint, elliptic
$\Psi$DOs, $A$ and $B$, on $M_D$. If we  assume that they commute,
then the following equality (the Wodzicki multiplicative formula) holds
\cite{kass89-177-199} \begin{eqnarray} a(A,B)=\frac{\mbox{res}\left[
(\ln(A^bB^{-a}))^2 \right]}{2ab(a+b)}=a(B,A)
\:,\label{wod3}\end{eqnarray} where $a >0$ and $ b> 0$ are the orders
of $A$ and $B$, respectively.

A sketch of the proof is presented in
what follows.
Recall that if $B$ is an elliptic operator of order $b>q$, according to
Wodzicki \cite{kass89-177-199}, one has the following property for the
non-commutative residue related to the $\Psi$DO $Q$ (Eq.~(\ref{wod1})
is just a consequence of it): in a neighborhood
of $z=0$, it holds \begin{eqnarray} \mbox{Tr}\,  
(QB^{-z})=\frac{\mbox{res}(Q)}{zb}+C \frac{\mbox{res}(Q)}{b}+r_Q(B)+O(z)
\:.\label{A1}\end{eqnarray}
Here $C$ is the Euler-Mascheroni constant and $r_Q(B)$ a coefficient 
depending also on $B$. As a consequence
\begin{eqnarray}
\mbox{Tr}\,   \left[ Q \left( B^{-\gamma_1z}-B^{-\gamma_2z}\right)\right]=
\frac{\left( \gamma_2-\gamma_1\right)}{zb 
\gamma_1\gamma_2} \mbox{res}(Q)+O(z)
\:,\label{r6}\end{eqnarray}
in which  and $\gamma_1$, $\gamma_2$ are positive real numbers.

Then it follows that, if $\eta$ is a $\Psi$DO of zero
order and $B$ a $\Psi$DO of positive order $b$,
and $x$ a positive real number,   then, in a neighborhood of $s=0$, one
has $\eta^{-xs}=1-xs \ln \eta+O(s^2)$ and 
\begin{eqnarray} s\mbox{Tr}\,   \left[\ln \eta
\eta^{-xs}(B^{-\gamma_1 s}-B^{-\gamma_2 s})\right]=\frac{(\gamma_2-\gamma_1)\mbox{res} (\ln
\eta)}{\gamma_1 \gamma_2 b}- sx \frac{(\gamma_2-\gamma_1)\mbox{res} 
[(\ln \eta)^2]}
{\gamma_2 \gamma_1 b}+O(s^2) \:.\label{a2}\end{eqnarray} As a consequence,
\begin{eqnarray} \lim_{s \to 0}  \partial_s \left[ s\mathop{\rm
Tr}  (\ln \eta \eta^{-xs}(B^{-\gamma_1 s}-B^{-\gamma_2 s})\right] =
 - x (\gamma_2-\gamma_1)\frac{\mbox{res} [(\ln
\eta)^2]}{\gamma_2 \gamma_1 b} \:.\label{a3}\end{eqnarray}

Consider  now two invertible, commuting, elliptic, self-adjoint
operators $A$ and $B$ on $M_D$, with $a$ and $b$ being the orders  of
$A$ and $B$, respectively. Within the zeta-function definition of the
determinants, consider the quantity \begin{eqnarray} F(A,B)=\frac{\det
(AB)}{(\det A)( \det B)}=e^{a(A,B)} \:.\label{a4}\end{eqnarray}
Introduce then the family of  $\Psi$DOs \begin{eqnarray} A(x)=\eta^x
B^{\frac{a}{b}}\,, \qquad \eta=A^{b}B^{-a} \,, \end{eqnarray} and
define the function \begin{eqnarray} F(A(x),B)=\frac{\det
(A(x)B)}{(\det A(x))( \det B)} \:.\label{a5}\end{eqnarray} We get
\begin{eqnarray} F(A(0),B)=\frac{\det B^{\frac{a+b}{b}}}{(\det
B^{\frac{a}{b}})( \det B)}=1 \,, \qquad F(A(\frac{1}{b}),B)=\frac{\det
(AB)}{(\det A)(\det B)}=F(A,B) \:.\label{a6}\end{eqnarray} As a
consequence, one  is led to deal with  the following expression for
the anomaly \begin{eqnarray} a(A(x),B)=\ln F(A(x),B)=-\lim_{s \to 0} \ 
\partial_s \left[ \mbox{Tr}\,   ( A(x)B)^{-s}-\mathop{\rm
Tr}  A(x)^{-s}-\mbox{Tr}\,   B^{-s} \right]
\:.\label{a7}\end{eqnarray} This quantity has the properties:
$a(A(0),B)=0$ and $a(A(\frac{1}{b}),B)= a(A,B)$.

The next step is to compute the first derivative of $a(A(x),B)$ with
respect to $x$, the result being \begin{eqnarray} \partial_x
a(A(x),B)=\lim_{s \to 0} \ \partial_s  s\left[ \mbox{Tr}\,  
\left( \ln \eta \eta^{-xs} \left( B^{-s\frac{a+b}{b}} -  B^{-s\frac{a}{b}}\right)
\right) \right] \:.\label{a8}\end{eqnarray} Making  use of Eq.~(\ref{a3}),
one  obtains \begin{eqnarray} \partial_x a(A(x),B)= x  \frac{b}{a(a+b)} \mbox{res}
[(\ln \eta)^2] \:.\label{a9}\end{eqnarray} And, finally, performing
the integration with respect to $x$,  from $0$ to $1/b$, one gets
Wodzicki's formula for the multiplicative anomaly, namely
\begin{eqnarray} a(A,B)=a(B,A)=\frac{\mbox{res}\left[
(\ln(A^bB^{-a}))^2 \right]}{2ab(a+b)} \:.\label{a10}\end{eqnarray} It
should be noted that $a(A,B)$ depends on a classical $\Psi$DO  of zero order.
Thus, it is independent on the renormalization scale $M$
appearing in the path integral.

We conclude this section by observing that the Wodzicki formula is also
valid for  $\Psi$DOs formally non self-adjoint, provided they are
complex  functions of self-adjoint elliptic operators. An example is
given by $K_\pm$.

\section{The multiplicative anomaly for the interacting $O(2)$ model}

In this section we come back to the problem of the  computation of the
multiplicative anomaly in the model considered in
Sec. 3. Putting $h_D=0$, which is proportional to the coupling constant,
we also get the results valid in the free case. Strictly speaking, the results of the last section 
are valid
for a compact manifold, but in the case of ${\hbox{{\rm I}\kern-.2em\hbox{\rm R}}}^D$, or in the finite
temperature case $S_1 \times {\hbox{{\rm I}\kern-.2em\hbox{\rm R}}}^{D-1}$, the divergence is trivial,
being contained in the spatial volume factor.

Since the multiplicative anomaly  depends on the regularization of the
ultraviolet divergences (in our approach this is equivalent to the
asymptotic behavior of the eingenvalues at infinity and to the related
divergence of the functional determinant) it follows that, due to the
fact that the Minkowski  space-time is ultrastatic ($g_{00}=-1$), its dependence
on the temperature is simply proportional to $\beta$, as we will see.

First, let us consider the self-adjoint factorization. Now the order 
of $L_\pm$ is $2$. Then, the Wodzicki
formula gives \begin{eqnarray} a_D(L_+,L_-)=\frac{1}{8} \,
\mbox{res}\left[ (\ln(L_+ L_-^{-1}))^2 \right]
\:.\label{wod446}\end{eqnarray} There is no ordering problem because 
we are dealing with commuting operators. We have to construct the complete
symbol $A(x,k)$ of the classical $\Psi$DO of zero order  $[\ln (L_+
L_-^{-1})]^2$. 
The complete symbol reads \begin{eqnarray}
A(x,k)&=&\left[ \ln \left(
k^2+m^2+e^2\mu^2+h_D+\left(\frac{4}{D^2}h^2_D+4e^2\mu^2 (\vec k^2
+m^2+h_D) \right)^{\frac{1}{2}} \right) \right. \nonumber \\ &-&\left.
\ln \left( k^2+m^2+e^2\mu^2+h_D-\left( \frac{4}{D^2}h^2_D+4e^2\mu^2
(\vec k^2 +m^2+h_D) \right)^{\frac{1}{2}} \right) \right]^2
\:,\label{902}\end{eqnarray} where $k^2= \vec{k}^2+k_\tau^2$.
Let us denote its asymptotics for large $|k|$ as \begin{eqnarray}
A(x,k) \equiv \sum_j A_{-j}(k) \:.\label{as}\end{eqnarray} According
to Wodzicki, we have to select the component $A_{-D}$ and use the 
multiplicative anomaly formula, namely
\begin{eqnarray} a_D(L_+,L_-)=\frac{1}{8}
\int_{M_D}\frac{dx}{(2\pi)^{D}}\int_{|k|=1}A_{-D}(k)dk
=\frac{1}{8}\frac{\beta V_N}{(2\pi)^{D}}\int_{|k|=1}A_{-D}(k)dk
\:.\label{wb}\end{eqnarray}  For
large $k^2$ and $k_\tau^2$, due to the $k^2$ dependence, it follows
that the  homogeneous components with odd indices are vanishing. As a
consequence, from Eq.~(\ref{wb}) one immediately gets the following:
for $D$ odd, the multiplicative anomaly vanishes.  This result is
consistent with the general theorem contained in \cite{kont95b}.

For $D$ even, the anomaly is present and the asymptotic expansion,
from which one can easily read off the  even homogeneous components, is
\begin{eqnarray} A(x,k)&=&\frac{1}{(k^2)^2} \left[ 16e^2\mu^2  \vec
k^2-32 (m^2+e^2\mu^2+h_D)\frac{e^2\mu^2 \vec
k^2}{k^2}+h_D^2+16e^2\mu^2(m^2+h_D) \right. \nonumber \\ &+& \left.
\frac{128}{3} \frac{e^4\mu^4 (\vec
k^2)^2}{(k^2)^2}+O(\frac{1}{k^2})\right] \:. \label{91}\end{eqnarray}
Thus, the first non vanishing even components are \begin{eqnarray}
A_{-2}(x,k)&=&\frac{16e^2\mu^2 \vec k^2}{(k^2)^2}, \nonumber \\
A_{-4}(x,k)&=&\frac{1}{(k^2)^2} \left[ -32
(m^2+e^2\mu^2+h_D)\frac{e^2\mu^2 \vec k^2}{k^2}+h_D^2+16e^2\mu^2(m^2+h_D)
\right.\nonumber \\ &+& \left. \frac{128}{3} \frac{e^4\mu^4 (\vec
k^2)^2}{(k^2)^2} \right] \:.\label{92}\end{eqnarray} It follows that
\begin{eqnarray} a_2(L_+,L_-)=\frac{\beta V_1}{2 \pi}e^2\mu^2
\:,\label{6}\end{eqnarray} and \begin{eqnarray} a_4(L_+,L_-)=
\frac{\beta V_3}{8\pi^2} \left[
e^2\mu^2(\frac{e^2\mu^2}{3}-m^2-h_D)+\frac{h_D^2}{8} \right]
\:.\label{7}\end{eqnarray} If $e=0$, we recover the results of Ref.
\cite{eliz97u-394}, in particular the absence of the multiplicative
anomaly for $D=2$. In the free case $h_D=0$ and we have \begin{eqnarray}
a_4(L_+,L_-)= \frac{\beta V_3}{8\pi^2} \left[ 
e^2\mu^2(\frac{e^2\mu^2}{3}-m^2) \right] \:.\label{77}\end{eqnarray}

Let us consider the other factorization. The complete symbol reads now
\begin{eqnarray} A(x,k)&=&\left[ \ln \left(
k^2+m^2-e^2\mu^2+h_D+\left( \frac{4}{D^2} h^2_D-4e^2\mu^2  k_\tau^2 
\right)^{\frac{1}{2}} \right) \right. \nonumber \\ &-&\left. \ln
\left( k^2+m^2-e^2\mu^2+h_D-\left(  \frac{4}{D^2} h^2_D-4e^2\mu^2 
k_\tau^2 \right)^{\frac{1}{2}} \right) \right]^2
\:,\label{903}\end{eqnarray} For large $k^2$ and $k_\tau^2$, due to
the $k^2$ dependence, it follows again that the  homogeneous
components with odd indices are vanishing and, for $D$ odd, the
multiplicative anomaly is absent as in the other case.

For $D$ even,  the first non vanishing even components are
\begin{eqnarray} A_{-2}(x,k)&=&-\frac{16e^2\mu^2 
k_\tau^2}{(k^2)^2},\nonumber  \\ A_{-4}(x,k)&=&\frac{1}{(k^2)^2}
\left[ 32 (m^2+h_D-e^2\mu^2)\frac{e^2\mu^2  k_\tau^2}{k^2}+h_D^2+
\frac{128}{3} \frac{e^4\mu^4 ( k_\tau^2)^2}{(k^2)^2} \right]
\:.\label{922}\end{eqnarray} A straightforward calculation gives
\begin{eqnarray} a_2(K_+,K_-)=-\frac{\beta V_1}{2 \pi}e^2\mu^2
\label{66}\end{eqnarray} and \begin{eqnarray}
a_4(K_+,K_-)=\frac{\beta V_3}{8\pi^2} \left[ e^2\mu^2(
m^2+h_D-\frac{e^2\mu^2}{3})+\frac{h_D^2}{8} \right]
\:.\label{78}\end{eqnarray} Again, for $e \to 0$ we get the result of
\cite{eliz97u-394}.

Some remarks are in order. According to the results of Sec. 2 and this
section, it seems quite natural to make the conjecture that, in the
free case and for $D$ even, one has \begin{eqnarray}
a_D(K_+,K_-)=-a_D(L_+,L_-) \:.\label{evena}\end{eqnarray} We have
proved this conjecture for $D=2$ and $D=4$, with an explicit
computation, namely \begin{eqnarray} a_2(K_+,K_-)=-\frac{\beta V_1}{2
\pi}e^2\mu^2\,, \qquad a_4(K_+,K_-)=\frac{\beta V_3}{8\pi^2} \left[
e^2\mu^2( m^2-\frac{e^2\mu^2}{3}) \right] \:.\label{ex1}\end{eqnarray}
Thus, in the free case, we also have \begin{eqnarray} \ln
Z_{\beta}(L_+,L_-)=\ln Z_{\beta}(K_+,K_-) \:,\label{nmm}\end{eqnarray}
On the other hand, assuming Eq.~(\ref{nmm}) and Eq.~(\ref{evena}) one
obtains a general expression for the multiplicative anomaly, namely
\begin{eqnarray} a_D(K_+,K_-)=-a_D(L_+,L_-)=-\beta V_N {\cal
E}_D(m,\mu) \:,\label{gma}\end{eqnarray} where ${\cal E}_D(m,\mu)$ is
given by Eq.~(\ref{calli}). For instance, for $D=6$, we have
\begin{eqnarray} a_6(K_+,K_-)&=&\frac{\beta V_5}{16\pi^3} \left[
e^2\mu^2( -\frac{1}{4}m^4+\frac{e^2\mu^2 m^2}{6}-\frac{2 e^4\mu^4
}{45}) \right] \:.\label{a66}\end{eqnarray} As a consequence, in the
free case, one obtains \begin{eqnarray} \ln
Z_{\beta,\mu}(K_+,K_-)=-\beta V_N {\cal E}_V(m,M)+V_N {\cal
S}(\beta,\mu)+\frac{1}{2}\beta V_N  {\cal E}_D(m,\mu)
\:,\label{free}\end{eqnarray} where \begin{eqnarray} {\cal E}_V
=\frac{1}{V_N} E_V\,, \qquad   {\cal S}(\beta,\mu)=
\frac{1}{V_N}S(\beta,\mu)\,. \end{eqnarray}

\section{The free charged bosonic model at finite temperature}

The presence of this new anomaly term in the grand canonical partition
 requires a reanalysis of the statistical mechanical
behaviour. This is better accomplished by the use of the effective
potential formalism at finite temperature and charge. Part of the
material of this section has appeared before (see for example,
\cite{bens91-44-2480} and the references quoted therein). We shall 
keep working in arbitrary spacetime dimensions, returning to 
the physically important $4D$ case at the end of this section.

In order to study the effective action in the presence of a non vanishing mean 
field, it is necessary to introduce the one-loop grand partition
function in the presence of external (constant) sources $J_i$ and 
the chemical potential $\mu$. This is done with the addition of
the coupling term $ -\int dx^D (J_i \phi_i)$ in the action. As the 
action is quadratic in the field degrees of freedom, a 
trivial Gaussian integration gives \begin{eqnarray}
Z_{\beta}(\mu,J_i)=\exp\left[\beta
V_N\frac{J_iJ_i}{2(m^2-e^2\mu^2)}\right]\,Z_{\beta}(\mu)
\:.\label{gpf}\end{eqnarray} where $V_N$ is the spatial volume. 
In terms of generating functionals per
unit volume \begin{eqnarray} V(\beta,\mu,J_i)=-\frac{1}{\beta V_N}\ln
Z_{\beta}(\mu,J_i) \:, \qquad V_0(\beta,\mu)=-\frac{1}{\beta V_N}\ln
Z_{\beta}(\mu) \:, \end{eqnarray} we have \begin{eqnarray}
V(\beta,\mu,J_i)=V_0(\beta,\mu)-\frac{ J_iJ_i}{2 (m^2-e^2\mu^2)}
\:.\label{gnf1}\end{eqnarray} which is nothing but the thermodynamics 
potential of the non interacting boson gas. 
The effective potential in presence of
external  sources is the Legendre transform of the generating
functional per unit volume (that is to say, the thermodynamics free 
energy density), and reads \begin{eqnarray} 
F(\beta,\rho,\Phi_i)=V(\beta,\mu,J_i)+\mu \rho+J_i
\Phi_i, \end{eqnarray} where $\mu$ and $J_i$ have to be expressed as 
functions of the charge density, 
$\rho$, and the mean field $\Phi_i$, by solving the defining equations
\begin{eqnarray} \rho=-\frac{\partial
V_\beta}{\partial \mu}=\frac{<Q>}{V_N}\,,\qquad \Phi_i=-
\frac{\partial  V_\beta}{\partial J_i}=\frac{1}{V_N}\int dx^N <\phi_i >
\:,\label{q}\end{eqnarray}
where $V_{\beta}$ is a shortened notation for $V(\beta,\mu,J_i)$. 
As a consequence, with $x=|\vec \Phi|$, eliminating
the external sources, one has \begin{eqnarray} 
F(\beta,\rho,x)=V_0(\beta,\mu)-\mu \frac{\partial
V_0(\beta,\mu)}{\partial \mu}+\frac{1}{2}(m^2+e^2\mu^2)x^2 \:,
\label{v1} \end{eqnarray} \begin{eqnarray} \rho=-\frac{\partial
V_0(\beta,\mu)}{\partial \mu} + e^2\mu x^2 \:,\label{v2}\end{eqnarray}
where $\mu$ is assumed to be a function $\mu=\mu(\beta,\rho,x)$, obtained by
solving Eq.~(\ref{v2}) in the unknown $\mu$.

The possible equilibrium states correspond to minima of the effective
potential with respect to $x$, at fixed temperature and charge density. Making
use of Eq.~(\ref{v1}) and Eq.~(\ref{v2}), one arrives at
\begin{eqnarray} \frac{\partial F}{\partial x}= x
(m^2-e^2\mu^2)\,,\qquad \frac{\partial^2 F}{\partial
x^2}= (m^2-e^2\mu^2)-2x e^2\mu \frac{\partial \mu}{\partial x}.
\end{eqnarray} Requiring the first derivative to vanish gives either:

(i) The unbroken phase, with solution $x=0$. From the expression of
the second derivative this will be a minimum provided
 $e\mu <m$, and the
related free energy density will be \begin{eqnarray} {\cal F}_\beta
&=&\mbox{min}_x\, F(\beta,\rho,x)=V_0(\beta,\mu)-\mu
\frac{\partial V_0(\beta,\mu)}{\partial \mu}= V_0(\beta,\mu)+\mu \rho\:,
\nonumber \\ \rho&=&-\frac{\partial V_0(\beta,\mu)}{\partial \mu}
\:.\label{f1}\end{eqnarray}

(ii) The symmetry breaking solution \begin{eqnarray} e\mu=\pm m
\:,\label{sbs}\end{eqnarray} here derived as an extremal property of
the effective potential. In this case $|\Phi|>0$, and assuming $e\mu=m$, 
one has \begin{eqnarray} {\cal F}_\beta
&=&\mbox{min}_x\, F(\beta,\rho,x)= V_0(\beta,e\mu=m)+\frac{m}{e}
\rho\:, \nonumber \\ emx^2&=&\rho+\left. \frac{\partial
V_0(\beta,\mu)}{\partial \mu}\right|_{e\mu=m}
\:.\label{f2}\end{eqnarray} In this symmetry breaking phase, one has
the phenomenon of Bose-Einstein condensation (BEC). From the last
equation and the fact that $V_0(\beta,\mu)$ is a monotonically 
decreasing function of $\beta$, it follows
 that $x^2=\Phi_i\Phi_i$ is non vanishing when $\beta>\beta_c$
, the inverse
of the critical temperature of the BEC transition, as given by the condition
\begin{eqnarray} \rho=-\left. \frac{\partial V_0(\beta_c,\mu)}{\partial
\mu}\right|_{e\mu=m} \:,\label{ct}\end{eqnarray} in agreement with the
recent criterium for the relativistic Bose-Einstein condensation
proposed in \cite{kirs96-368-119,kirs97u-161}. For $\beta\leq\beta_c$, 
on the other hand, $|\Phi|=0$ and the $U(1)$-symmetry is restored.

Now in the free case we know
$V_\beta$. In the unbroken phase, $x=0$, $e|\mu| <m$, $\beta
<\beta_c$, and including the anomaly term, Eq.~(\ref{calli}), with
$2Q=D$, one has \begin{eqnarray} {\cal F}_\beta&=&{\cal
E}_V-\frac{4 m^{\frac{D}{2}}}{(2\pi)^{\frac{D}{2}}}\sum_{r=1}^\infty
\left(\frac{1}{\beta r}\right)^{\frac{D}{2}-1}\cosh (r e \mu \beta)
K_{\frac{D}{2}}(r m \beta)+\mu \rho+\frac{1}{2}\sum_{r=1}^{Q}c_{Q,r} (e\mu)^{2r}
m^{2Q-2r}\, ,\nonumber \\ \rho &=&\frac{4 e
m^{\frac{D}{2}}}{(2\pi)^{\frac{D}{2}}}\sum_{r=1}^\infty
\left(\frac{1}{\beta r}\right)^{\frac{D}{2}-1}\sinh (r e \mu \beta)
K_{\frac{D}{2}}(r m \beta)-  e\sum_{r=1}^{Q}c_{Q,r} r
(e\mu)^{2r-1}m^{2Q-2r} \:.\label{ct11} \end{eqnarray}
The last term in both equations accounts for the anomaly.

In the broken phase, $e|\mu|=m$, $\beta > \beta_c$, one obtains
\begin{eqnarray} {\cal F}_\beta&=&{\cal E}_V-\frac{4
m^{\frac{D}{2}}}{(2\pi)^{\frac{D}{2}}}\sum_{r=1}^\infty
\left(\frac{1}{\beta r}\right)^{\frac{D}{2}-1}\cosh (r m \beta)
K_{\frac{D}{2}}(r m \beta)+\frac{1}{2}\sum_{r=1}^{Q}c_{Q,r}
m^{2Q}+\frac{m}{e} \rho\, ,\nonumber \\ \rho &=&\frac{4 e
m^{\frac{D}{2}}}{(2\pi)^{\frac{D}{2}}}\sum_{r=1}^\infty
\left(\frac{1}{\beta r}\right)^{\frac{D}{2}-1}\sinh (r m \beta)
K_{\frac{D}{2}}(r m \beta)-e\sum_{r=1}^{Q}r c_{Q,r} m^{2Q-1}
+e m x^2 \:.\label{ct111} \end{eqnarray} In this case, if we formally
take the zero temperature limit $ \beta \to \infty$, we get
\begin{eqnarray} \rho \to - e \sum_{r=1}^{Q}r c_{Q,r}
m^{2Q-1}+e mx^2 \:.\label{bbb}\end{eqnarray} Now observe that the
anomaly contribution leads to some problems, since it does not have a
definite sign. For example, for any fixed $\rho =0$, $x^2$ in 
(\ref{bbb}) could be made negative when the anomaly
contribution is negative. 
We thus arrive at a contradiction, as happens, for example, for $D=6$.

At this point, however, we should notice that our
discussion has involved ``unrenormalized" quantities only, regularized by using
the zeta-function procedure. In particular, the charge operator has not 
been properly defined, and the above result indicates that this operator 
was not normal ordered relative to the  Minkowski vacuum. Now the 
charge operator appears in the Hamiltonian multiplied by $\mu$. 
Therefore any normal ordering ambiguity in the charge operator gives 
rise to an ambiguity in the effective action, which must be a linear
homogeneous function of $\mu$, namely $K\mu$, with a freely 
disposable constant $K$. We have therefore the freedom to define the 
generating functional $V(\beta,\mu,J_i)$ up to the linear term $K\mu$.
This will not change the effective action, the Legendre transform being 
unaffected by linear terms. We then introduce the renormalized generating
functional \begin{eqnarray}
V^R=V+ \mu K={\cal E}_V +\frac{1}{2}{\cal E}_D(m,\mu)- \frac{{\cal
S}(\beta,\mu)}{\beta}-\frac{ J_iJ_i}{2 (m^2-e^2\mu^2)}+
\mu K, \end{eqnarray}
where ${\cal E}_D(m,\mu)$ is the anomaly, given by Eq.~(\ref{calli}) 
and ${\cal S}(\beta,\mu)$ was defined in Eq.~(\ref{ss1}). 
So we have also a finite 
renormalization of the charge density: \begin{eqnarray} \rho^R=-
\frac{\partial V_0^R}{\partial \mu} +e\mu x^2=\rho-K. \end{eqnarray}
Since the effective potential is the same, the extremals are unchanged
and we may rewrite the free energy in the symmetry breaking phase as
\begin{eqnarray} {\cal F}_\beta &=&{\cal E}_V-\frac{1}{\beta} {\cal
S}(\beta,e\mu=m)+\frac{m}{e} \rho^R+ \frac{m}{e} K +\frac{1}{2}
\sum_{r=1}^{Q}c_{Q,r} m^{2Q}, \nonumber \\ 
\rho^R&=&-\frac{1}{\beta}\frac{\partial {\cal S}(\beta,\mu)}{\partial
\mu}|_{e\mu=m} -K-\frac{e}{m} \sum_{r=1}^{Q}r c_{Q,r} m^{2Q} +e m x^2
\:.\label{sbp1}\end{eqnarray} On the
other hand, in the unbroken symmetric phase we have,  \begin{eqnarray} {\cal F}_\beta &=&{\cal
E}_V-\frac{1}{\beta} {\cal S}(\beta,\mu)+\mu \rho^R+ \mu K +\frac{1}{2}
\sum_{r=1}^{Q}c_{Q,r} (e\mu)^{2r} m^{2Q-2r}, \nonumber \\ 
\rho^R&=&-\frac{1}{\beta}\frac{\partial {\cal S}(\beta,\mu)}{\partial
\mu} -K-e\sum_{r=1}^{Q}r c_{Q,r}  (e\mu)^{2r-1}m^{2Q-2r}
\:.\label{sbp111}\end{eqnarray}

The constant $K$ can now be fixed by imposing normalization 
conditions. By doing this we see how the spacetime dimension
affects renormalization and how special the case $D=4$ is.
We do so just by demanding that, at zero temperature and
charge density, the symmetry be unbroken, which sounds as a very natural 
normalization condition. Using (\ref{sbp1}), this fixes $K$ to be
\begin{eqnarray} K= -e\sum_{r=1}^{Q}rc_{Q,r} m^{2Q-1}
\:.\label{bb}\end{eqnarray} 
As a result, in the broken phase $(|e\mu|=m)$ 
one obtains \begin{eqnarray} {\cal F}_\beta &=&{\cal E}_V-\frac{1}{\beta}
{\cal S}(\beta,e\mu=m)+\frac{m}{e} \rho^R+\frac{1}{2}
\sum_{r=1}^{Q}c_{Q,r} (1-2 r) m^{2Q}, \nonumber \\ 
\rho^R&=&-\frac{1}{\beta}\frac{\partial {\cal S}(\beta,\mu)}{\partial
\mu}|_{e\mu=m}+e m x^2 \:,\label{asbp111}\end{eqnarray} while, in the
symmetric phase, one gets \begin{eqnarray} {\cal F}_\beta &=&{\cal
E}_V-\frac{1}{\beta} {\cal S}(\beta,\mu)+\mu \rho^R- e \mu
\sum_{r=1}^{Q}rc_{Q,r} m^{D-1}+\frac{1}{2}\sum_{r=1}^{Q}c_{Q,r}
(e\mu)^{2r} m^{2Q-2r}, \nonumber \\ 
\rho^R&=&-\frac{1}{\beta}\frac{\partial {\cal S}(\beta,\mu)}{\partial
\mu}+e\sum_{r=1}^{Q}rc_{Q,r} m^{2Q-1}-e
\sum_{r=1}^{Q}r c_{Q,r}  (e\mu)^{2r-1}m^{2Q-2r}
\:.\label{sbp1112}\end{eqnarray} Note that in this symmetric phase,
one can now take the limit $e \to 0$ to reach the usual expression of
the free energy density for an uncharged boson gas. For general 
$D=2Q$, however, the free energy receives contributions from the anomaly 
term in both phases. On the other hand, with 
the above charge renormalization, the critical temperature is
given implicitly by (choosing $\mu >0 $) \begin{eqnarray}
\rho^R+\left.\frac{1}{\beta}\frac{\partial {\cal S}(\beta_c,\mu)}{\partial
\mu}\right|_{e\mu=m}=0 \:.\label{ccc}\end{eqnarray} which is the usual
condition for the critical temperature
\cite{kirs96-368-119,kirs97u-161}, and it remains unaffected by the anomaly.
It is important to recognize that 
the normalization condition just discussed is not the only one that is 
possible. Another physically natural condition would be to demand that 
at zero coupling, the free energy reduces to the free energy of 
uncharged bosons. This then fixes $K$ to be
\begin{eqnarray} K= -\frac{e}{2}\sum_{r=1}^{Q}c_{Q,r} m^{2Q-1}
\:.\label{bb1}\end{eqnarray} Now it is the free energy in the broken 
phase that remains unaffected, while $\beta_c$ changes. 

Let us consider now the physically  important case $D=4$. Then, from 
the expression (\ref{ex}) for the anomaly, we have in addition the 
remarkable identity
\begin{eqnarray} \sum_{r=1}^{2}r c_{2,r} = \frac{1}{2}
\sum_{r=1}^{2}c_{2,r}=\frac{1}{24\pi^2}
 \:.\label{bbb1}\end{eqnarray} This shows that the two normalization 
conditions just discussed are actually equivalent and fix a unique
value  $K=-\frac{e m^3}{24\pi^2}$. As a result, in the
symmetry breaking phase and for $D=4$, we obtain \begin{eqnarray}
{\cal F}_\beta &=&{\cal E}_V-\frac{1}{\beta} {\cal
S}(\beta,m)+\frac{m}{e}  \rho^R, \nonumber \\ e x^2&=&
\left. \frac{1}{m}\left( \rho^R+\frac{1}{\beta}\frac{\partial {\cal
S}(\beta,\mu)}{\partial \mu}\right|_{e\mu=m} \right)
\:.\label{sbp2}\end{eqnarray} which means that the broken phase is 
totally unaffected by the anomaly. 
On the other hand, in the symmetric phase $ \beta < \beta_C$,
$x=0$ and $e \mu < m$, the result
is \begin{eqnarray} {\cal F}_\beta &=&{\cal E}_V-\frac{1}{\beta} {\cal
S}(\beta,\mu)+\mu \rho^R -\frac{\mu e m^3}{12 \pi^2}+\frac{1}{8
\pi^2}\left( e^2\mu^2m^2-\frac{1}{3}e^4\mu^4 \right), \nonumber \\
\rho^R&=&-\frac{1}{\beta}\frac{\partial \cal{S(\beta,\mu)}}{\partial
\mu} -\frac{1}{8 \pi^2}\left( 2e^2\mu m^2-\frac{4}{3}e^4\mu^3 \right)
+\frac{em^3}{12\pi^2}\:. \label{sspp}\end{eqnarray} We observe that 
the anomaly term  (the term in square brackets in (\ref{sspp})) is present 
in the free energy above $T_C$. Moreover, it is   again vanishing as
 $e \to 0$, and the correct expression of the free energy
density for the uncharged boson gas is recovered. Although the 
multiplicative anomaly term itself 
may  be not negligible, 
as a numerical study of it shows, the 
renormalization used in order to preserve the vacuum structure has rendered the 
anomaly term quite harmless near $T_C$ and below. On the other hand, at 
ultra relativistic temperatures, $T\gg m$, it is negligible in 
comparison with the Planckian $T^4$-term, since the anomaly term increases with 
mass precisely as $m^4$ (however this holds whenever the charge 
density is much less than $T^3e^{-6}$, in order to neglect the Coulomb 
interaction). Our conclusion is that it gives a relevant correction at
intermediate temperatures, say of the order of $T\simeq m$, which are
relativistic for any known massive elementary particle.

\section{ Concluding remarks}

In this paper some deep results of the zeta-function regularization
procedure have been employed in order to study rigorously the one-loop
effective potential for a fixed charged self-interacting scalar field
at finite temperature. The chemical potential has been introduced
according to the  path integral approach
\cite{kapu81-24-426,bens91-44-2480} making use of method I, as
discussed recently in \cite{fili97u-90}. Method I leads to the
presence of a multiplicative anomaly term, whose existence had been
completely overlooked in the literature. Using powerful mathematical
expressions, this anomaly term can be computed exactly. Its explicit
form has been obtained here for $D=2$ and $D=4$ (while a simple
program yields it for {\it any} desired value of $D$.) We have shown
it to be vanishing for $D$ odd and also its fundamental importance in
getting factorization invariance (in the operational sense) of the
regularized one-loop effective action. On the other hand, assuming 
factorization invariance, we have obtained, in the free case, a
general expression for the multiplicative anomaly, valid for any even 
$D$.

The existence of this new  contribution has led us to revisit
and discuss the non-interacting case in detail. In particular, we have
reexamined the spontaneous symmetry breaking issue and the related
relativistic Bose-Einstein condensation phenomenon. A renormalization
of the charge density has been introduced in arbitrary even dimension
and has led to the usual expression for the critical temperature. In
particular, in the broken phase,  we have shown that only in the
physically important case $D=4$ the new contribution due to the 
multiplicative anomaly may be
absorbed in the charge renormalization process. However, in the
symmetric unbroken phase, the multiplicative anomaly term  gives a non
vanishing contribution, which has been overlooked in  previous 
investigations. Although it is non leading in the 
ultra high temperature regime, nevertheless it can give relevant correction at 
intermediate temperatures  of order $T\simeq m$, $m$ being the mass of the
charged boson.

In the interacting case, even in at the one-loop approximation, it is
quite difficult ---within the first factorization--- to deal with  the
closed expression for the zeta-functions. In fact, for $D=4$, we have
\begin{eqnarray}
\zeta(s|L_\pm)&=&\frac{\beta\Gamma(s-\frac{1}{2})}{2\sqrt \pi\Gamma(s)}
\zeta(s-\frac{1}{2}|L_3+e^2\mu^2+h \pm
\sqrt{\frac{h^2}{4}+4e^2\mu^2(L_3+h)})\nonumber \\ &-&2s \mathop{\rm
Tr}  \ln \left( 1-\exp{\beta \sqrt{L_3+e^2\mu^2+h\pm
\sqrt{\frac{h^2}{4}+4e^2\mu^2(L_3+h)}}}  \right)
+O(s^2)\:,\label{ffz}\end{eqnarray} while,  within the second
factorization, one has difficulties with the sum over the Matsubara
frequencies $\omega_n$. For example, again for $D=4$, one should deal
with \begin{eqnarray} \zeta(s|
K_\pm)=\frac{V_3}{\Gamma(s)}\int_0^\infty dt t^{s-1}\frac{
e^{-t(m^2+h)}}{(4\pi t)^{\frac{3}{2}}} \sum_n \exp\left[-t \left(
\omega^2_n-e^2\mu^2 \pm \sqrt{\frac{h^2}{4}-4e^2\mu^2
\omega^2_n}\right)\right] \:.\label{fff}\end{eqnarray} As a consequence 
---as
far as the factorization invariance of the partition function is
concerned---  the relevance
of the multiplicative anomaly (evaluated in Sec. 4)
appears manifest, in the interacting case. The issue will
require further investigation.

\section*{Acknowledgments}{We would like to thank Klaus Kirsten and
Guido Cognola  for valuable discussions. This work has been supported
by the cooperative agreement INFN (Italy)--DGICYT (Spain). EE has been
partly financed by DGICYT (Spain), projects PB93-0035 and PB96-0925,
and by  CIRIT
(Generalitat de Catalunya),  grant 1995SGR-00602. AF wishes to
acknowledge financial support from the European Commission under TMR
contract N. ERBFMBICT972020 and, previously, from the Foundation
BLANCEFLOR Boncompagni--Ludovisi, n\'ee Bildt. }

\section{Appendix}

In this Appendix we will show that the multiplicative anomaly is
present also in a large class of regularizations of functional
determinants appearing in the one-loop effective action. They are often
used in the literature and  can be called ``generalized proper-time
regularizations", since the prototype among them was introduced by
Schwinger \cite{schw51-82-664}. We will show that the presence of the
multiplicative anomaly stems from the obvious ---but crucial--- fact
that the finite part of all these regularization contains,
unavoidably, the zeta-function regularization contribution
$\zeta'(0|L)$.

If we work in the Euclidean formalism, the class of regularization we
shall be dealing with is defined by \cite{byts96-266-1,ball89-182-1}
\begin{eqnarray} \ln \det \left(
\frac{L}{M^2}\right)(g_\varepsilon(t))=-\int_0^\infty dt\ 
t^{-1}g_\varepsilon(t) \mbox{Tr}\,   e^{-t\frac{L}{M^2}}
\:,\label{s1}\end{eqnarray} where $g_\varepsilon(t)$ is a suitable
regularizing function, necessary in order to control, for $\varepsilon
>0$, the (ultraviolet) divergences for small $t$ in the integral and
$M^2$ is the mass renormalization parameter which renders $t$
adimensional. $L$ is an elliptic operator. We note that the
zeta-function regularization and the related Dowker-Critchley one
\cite{dowk76-13-3224} belong to this class. They correspond to
the choices \begin{eqnarray} g_\varepsilon^1(t)=\frac{d}{d
\varepsilon}\left( \frac{t^\varepsilon}{\Gamma(\varepsilon)} \right)
\,, \qquad g_\varepsilon^2(t)=t^\varepsilon \:.\label{s2}\end{eqnarray}
As a consequence, the first regularization, as is well known, gives a
finite result $\zeta'(\varepsilon|L)$, after an analytical
continuation in $\varepsilon$, when the parameter goes to zero, while
the second one gives ($C$ being again the Euler-Mascheroni constant)
\begin{eqnarray} \ln \det \left(
\frac{L}{M^2}\right)(g_\varepsilon^2(t))=-\Gamma(\varepsilon)\zeta(\varepsilon|\frac{L}{M^2})=
-\frac{1}{\varepsilon}\zeta(0|L)-\zeta'(0|\frac{L}{M^2}) +C
\zeta(0|L)+O(\varepsilon) \:.\label{s3}\end{eqnarray} In this case,
besides the divergent contribution, the finite part contains the
zeta-function regularization result.

Let us consider another class of proper-time regularizations such that
the functions $g_\varepsilon(t)$ admit the Mellin transform
\begin{eqnarray} \hat g_\varepsilon(s)=\int_0^\infty dt \ 
t^{s-1}g_\varepsilon(t) \:.\label{s4}\end{eqnarray} Three popular
examples are the ultraviolet cutoff regularization
$g_\varepsilon^3(t)=\theta(t-\varepsilon)$ \begin{eqnarray} \ln \det
\left(
\frac{L}{M^2}\right)(g^3_\varepsilon(t))=-\int_\varepsilon^\infty dt\ 
t^{-1} \mbox{Tr}\,   e^{-t\frac{L}{M^2}}\,,\qquad  \hat
g_\varepsilon^3(s)=-\frac{1}{s}\varepsilon^s
\:,\label{s44}\end{eqnarray} the point-splitting regularization
$g_\varepsilon^4(t)= e^{-\frac{\varepsilon}{t}}$ \begin{eqnarray} \ln
\det \left( \frac{L}{M^2}\right)(g^4_\varepsilon(t))=-\int_0^\infty
dt\ t^{-1}  e^{-\frac{\varepsilon}{t}}\mbox{Tr}\,  
e^{-t\frac{L}{M^2}} \,, \qquad\hat
g^4_\varepsilon(s)=\Gamma(-s)\varepsilon^s \:,\label{s5}\end{eqnarray}
and Pauli-Villars regularization, which in our formalism may be
expressed as \begin{eqnarray} g_\varepsilon^5(t)=\left(
1-e^{-t\frac{\alpha}{\varepsilon}} \right)^{D-1}
\:.\label{pv}\end{eqnarray} Thus \begin{eqnarray} \ln \det \left(
\frac{L}{M^2}\right)(g^5_\varepsilon(t))=-\int_0^\infty dt t^{-1}
\left( 1-e^{-t\frac{\alpha}{\varepsilon}} \right)^{D-1}  \mathop{\rm
Tr}  e^{-t\frac{L}{M^2}} \:,\label{s6}\end{eqnarray} with
\begin{eqnarray} \hat g^5_\varepsilon(s)=\frac{1}{s}+\sum_j
c_j(\varepsilon)\frac{1}{D-1+j+s} \:,\label{pv2}\end{eqnarray} where
$c_j(\varepsilon)$ are  constants which diverge in the limit
$\varepsilon \to 0 $. Making use of the Parceval-Mellin identity, we
may rewrite Eq.~(\ref{s1}) in terms of a complex integral involving
the Mellin transform of $ \mbox{Tr}\,  
e^{-t\frac{L}{M^2}}$, namely $\zeta(z|L)M^{2z}$ and
$g_\varepsilon(t)$, i.e. \begin{eqnarray} \ln \det \left(
\frac{L}{M^2}\right)(g_\varepsilon(t))=-\frac{1}{2 \pi
i}\int_{\mathop{\rm Re}   z>D/2} dz \ 
\Gamma(z)\zeta(z|L)M^{2z} \hat {g}_\varepsilon(-z) \:.\label{s61}
\end{eqnarray} Now the meromorphic properties of $\zeta(z|L)$ are
known and we have \begin{eqnarray}
\zeta(z|L)=\frac{1}{\Gamma(z)}\left( \sum_r
\frac{A_r}{z+r-D/2}+J(z)\right) \:,\label{mprop}\end{eqnarray} where 
$A_r$ are the Seeley-De Witt coefficients (computable) and $J(z)$ is
the analytical part (normally unknown). We also have
$\zeta(0|L)=A_{D/2}$, with $A_{D/2}=0$ when $D$ is odd, since we are
working in manifolds without boundary. Shifting the vertical contour
to the left, assuming that the Mellin transform $\hat
g_\varepsilon(-z)$ is regular for $\mathop{\rm Re}   z>0$ and
has a simple pole at $z=0$ given, with residue $1$ (see, for example,
Eqs.~(\ref{s44}) and (\ref{s5})), the residue theorem yields
\begin{eqnarray} \ln \det \left(
\frac{L}{M^2}\right)(g_\varepsilon(t))&=&-\zeta'(0|L)+\zeta(0|L)\left(
\ln M^2 +C \right) \nonumber \\ &-& \sum_{r\neq \frac{D}{2}} A_r
\hat {g}_\varepsilon(r-\frac{D}{2}) M^{D-\frac{r}{2}}-\zeta(0|L) \ln
\varepsilon+O(\varepsilon) \:.\label{anfor}\end{eqnarray} As a
consequence, the finite part, coming from the singularity at $z=0$
contains $\zeta'(0|L)$ and the multiplicative anomaly is also present
in this large class of regularizations. The sum over $r$ contains the
ultraviolet divergences, controlled by $\varepsilon$. In the one-loop
approximation, these divergent terms are removed by the corresponding
counterterms.

\end{document}